\documentstyle[prd,aps,epsf,eqsecnum]{revtex}
\begin{document}
\def\CC{{\Bbb C}}
\def\NN{{\Bbb N}}
\def\QQ{{\Bbb Q}}
\def\RR{{\Bbb R}}
\def\ZZ{{\Bbb Z}}
\def\cA{{\cal A}}          \def\cB{{\cal B}}          \def\cC{{\cal C}}
\def\cD{{\cal D}}          \def\cE{{\cal E}}          \def\cF{{\cal F}}
\def\cG{{\cal G}}          \def\cH{{\cal H}}          \def\cI{{\cal I}}
\def\cJ{{\cal J}}          \def\cK{{\cal K}}          \def\cL{{\cal L}} 
\def\cM{{\cal M}}          \def\cN{{\cal N}}          \def\cO{{\cal O}}
\def\cP{{\cal P}}          \def\cQ{{\cal Q}}          \def\cR{{\cal R}} 
\def\cS{{\cal S}}          \def\cT{{\cal T}}          \def\cU{{\cal U}}
\def\cV{{\cal V}}          \def\cW{{\cal W}}          \def\cX{{\cal X}}
\def\cY{{\cal Y}}          \def\cZ{{\cal Z}}

\def\arsinh{\mathop{\rm arsinh}\nolimits}
\def\arsinh{\mathop{\rm arcosh}\nolimits}
\def\arsin{\mathop{\rm arsin}\nolimits}
\newtheorem{theorem}{Theorem}
\newtheorem{prop}{Proposition}
\newtheorem{conj}{Conjecture}

\title{BRANE WORLD COSMOLOGY WITH GAUSS-BONNET INTERACTION}

\author{B. ABDESSELAM$^{\S,}$\footnote{Permanent address: 
Laboratoire de Physique Th\'eorique, Centre Universitaire Mustapha 
Stambouli, 29000 Mascara, Algeria.}$^{,}$\footnote{E-mail: 
boucif@celfi.phys.univ-tours.fr} and N. MOHAMMEDI$^{\S,}$\footnote{E-mail: 
nouri@celfi.phys.univ-tours.fr}}

\address{\textit{$^{\S}$
Laboratoire de Math\'ematiques et Physique Th\'eorique, 
Universit\'e Fran\c{c}ois Rabelais,\\
Facult\'e des Sciences et Techniques,  
Parc de Grandmont, F-37200 Tours, France.}}

\maketitle

\vspace{5mm}

\begin{abstract}
We study a Randall-Sundrum model modified by a Gauss-Bonnet interaction term. 
We consider, in particular, a Friedmann-Robertson-Walker metric on the brane
and analyse the resulting cosmological scenario. It is shown that the usual 
Friedmann equations are recovered on the brane. The equation of state relating 
the enery density and the pressure is uniquely determined by the matching conditions.
A cosmological solution with negative pressure is found.
    
\end{abstract}

\smallskip
\smallskip

\centerline{October 2001}



\section{Introduction}

The possibility that our universe is a four dimensional brane embedded in a 
higher dimensional spacetime has been extensively discussed recently. The most 
popular model in this context is the one proposed by Randall and Sundrum 
\cite{rs}.
This scenario is based on the metric
\begin{eqnarray}
ds^2 = A^2\left(y\right)\eta_{ij}dx^i dx^j + dy^2 \,\,\,,
\label{flatmetric}
\end{eqnarray}
where $\eta_{ij}$ is a flat Minkowski four dimensional metric on the brane
and $A^2(y)$ is the warp factor depending only on $|y|$.  
Perturbations of this metric reproduces 
the expected $1/r$ Newtonian potential on the brane (the observed universe). 
This is due to the fact 
that the zero modes of the perturbation propagate on the brane only (they tend 
rapidely to zero in the fifth dimension). The other modes (the massive 
Kaluza-Klein modes) give merely a correction in $1/r^3$ to this potential.
\par
One of the first development of this model was the generalisation of the 
Randall-Sundrum ansatz to include a wider class of metrics 
\cite{w1,w2,w3,w4,w5,w6,w7,w8,w9,w10,w11,w12,w13,w14}. Different 
geometries were treated by considering solutions to the Randall-Sundrum model 
with metrics which, up to a non-constant conformal factor, can be written as 
\begin{eqnarray}
ds^2 = A^2\left(y\right) g_{ij}\left(x,y\right)dx^i dx^j + dy^2 \,\,\,.
\label{curvedmetric}
\end{eqnarray}  
The requirement that the zero modes of the perturbations around these metrics 
are localised on the brane imposes further constraints on this class of 
geometries \cite{w2}.
\par
Since the Randall-Sundrum model is a string inspired picture\cite{witten}, 
one would like to understand the implications of higher curvature terms 
in such brane world 
universe. These terms naturally arise in the string effective action beyond 
the first order in the string tension $\alpha'$. The inclusion of these terms 
are also of relevance to cosmology and inflation.  It turns out that generic 
higher curvature terms lead in general to a delocalisation of gravity from the 
brane \cite{gb1,gb2}. A combination of these terms in a Gauss-Bonnet form yields, however, 
the desired Randall-Sundrum behaviour of the zero modes of the perturbations.
We should mention that the analyses of refs.\cite{gb1,gb2} is carried out with a brane 
possessing a flat metric of the form given in (\ref{flatmetric}). 
In this context, various other issues were also treated using higher curvature
terms \cite{hg1,hg3,hg4,hg5,hg6,hg7,hg8,hg9,hg10}.   
\par
In this letter, we deal with a Randall-Sundrum model complemented by a 
Gauss-Bonnet density where the five dimensional metric is of the form in 
(\ref{curvedmetric}). We start by considering a metric on the brane with 
spherical symmetry. It is shown that the only possible solution in this 
case is a de Sitter or anti de Sitter spacetime for the brane.  
This is in contrast to the case without a Gauss-Bonnet term where 
black hole geometries are allowed \cite{w2}. Our bulk metric involves a warp factor that 
presents an oscillatory regime among other possibilities.
\par 
A second study consists in taking a Friedmann-Robertson-Walker
metric on the brane. We recover the equations of ordinary cosmology
on the brane. This is to be compared to previous brane world cosmology 
models 
\cite{bc1,bc2,bc3,bc4,bc5,bc6,bc7,bc8,bc9,bc10,bc11,bc12,bc13,bc14,bc15,bc16,bc17,bc18,bc19,bc20,bc21}
where, among other things, 
the square of the Hubble parameter is found to be proportional to the
square of the energy density. 
Here, the matching conditions are so restrictive that
they determine the equation of state relating the energy density to the
pressure. Various inflationary solutions with a cosmological constant
on the brane are determined. Another solution with a time dependent energy density and 
pressure is also found. However, the pressure for this matter is negative and cannot
describe an ordinary dust. On the other hand, a scalar field is found 
whose energy-momentum tensor could describe this behaviour.

\section{The Model}
We consider a five dimensional spacetime with coordinates $(x^0\equiv t,
x^1,x^2,x^3,x^4\equiv y)$ where $(t,x^1,x^2,x^3)$ denotes the usual 
four-dimensional spacetime and $x^4\equiv y$ is the coordinate of the fifth 
dimension, which is an orbifold $S^1/Z_2$ where the $Z_2$ action identifies 
$y$ and $-y$. The five dimensional indices are denoted by $M,\;N,\dots=0,\dots,4$ and 
the four dimensional brane world indices are $i,\;j,\dots=0,\dots,3$. 
We will neglect the matter interaction 
and consider the five dimensional gravitational action
\begin{eqnarray}
S &=& \int d^5x\sqrt{-G}\left(\alpha R+\Lambda + \beta L_{{\rm GB}} \right)
\end{eqnarray}
where $\alpha$ and $\beta$ are two coupling constants and $\Lambda$ is 
the five dimensional cosmological constant. The Gauss-Bonnet Lagrangian density
is 
\begin{eqnarray}
L_{{\rm GB}}= R_{MNPQ}R^{MNPQ}-4R_{MN}R^{MN}+R^2 \,\,\,.
\end{eqnarray}
The equations of motion corresponding to our action are\footnote{Our conventions are
such that $R^M_{\,\,\,NPQ}=\partial_P\Gamma^M_{NQ} + \Gamma^M_{PR}\Gamma^R_{NQ} -
\left(P\leftrightarrow Q\right)$ and $R_{MN}=R^Q_{\,\,\,MQN}$.} 
\begin{eqnarray}
{\cal E}_{MN} &\equiv& 
\alpha\left(R_{MN}-{1\over 2}g_{MN}R \right) -{1\over 2}\Lambda g_{MN}
\nonumber\\
&+&2\beta\biggl(R_{M}^{\,\,\,PQS}R_{NPQS}+2R^{PQ}R_{MPQN}-2R_{MP}R_{N}^{\,\,\,P}
+RR_{MN} -{1\over 4}g_{MN}L_{{\rm GB}}\biggr)=0\,\,\,. 
\end{eqnarray}
As it is well-known, these equations possess the following solutions \cite{gbs1,gbs2}
\begin{eqnarray}
R_{MNPQ}=-\sigma\left(G_{MP}G_{NQ}- G_{MQ}G_{NP}\right) \,\,\,,
\end{eqnarray}
where $\sigma$ takes two possible values as given by
\begin{eqnarray}
\sigma={1\over 4\beta}\left(\alpha  \pm \sqrt{\alpha^2 - {2\over 3}\beta\Lambda}\right)
\,\,\,.
\label{sigma}
\end{eqnarray}
This last equation can be inverted to get an expression for $\Lambda$
\begin{eqnarray}
\Lambda= 12\sigma\left(\alpha-2\beta\sigma\right)
\,\,\,.
\label{Lambda}
\end{eqnarray}
This value of $\Lambda$ will be needed in the rest of the paper.
\par
As a start, we would like to analyse the above equations of motion in the context of the brane 
world scenario. We consider, for this purpose, a spherically symmetric line element as given by
\begin{eqnarray}
ds^2=A^2(y)\biggl(- N(r)dt^2+M(r)dr^2+r^2(d\theta^2+
\sin^2\theta d\phi^2)\biggr)+dy^2\,\,\,.
\end{eqnarray}
and examin the equations of motion
\begin{eqnarray}
{\cal H}^M_N \equiv {\cal E}^M_N+T^M_N=0\,\,\,\,,
\end{eqnarray}
where the non-zero components of the energy-momentum tensor $T^M_N$ are 
\begin{eqnarray}
T^i_j=\delta^i_j \lambda\delta\left(y\right) 
\,\,\, 
\end{eqnarray}
with $\lambda$ denoting the cosmological constant on the brane. 
We obtain four different equations corresponding to the components ${\cal H}^0_0$, 
${\cal H}^1_1$, ${\cal H}^2_2={\cal H}^3_3$ and ${\cal H}^4_4$. Substracting ${\cal H}^1_1$
from ${\cal H}^0_0$ leads to
\begin{eqnarray}
M\left(r\right)={1 \over N\left(r\right)}\,\,\,.
\end{eqnarray}
Substituting for $M\left(r\right)$ and substracting ${\cal H}^2_2$ from ${\cal H}^0_0$
yields
\begin{eqnarray}
N\left(r\right)=1+\mu r^2 +{\nu\over r}
\,\,\,,
\end{eqnarray}
where $\mu$ and $\nu$ are two constants of integration. 
\par
After substituting for $N\left(r\right)$, the equation corresponding to ${\cal H}^4_4=0$
can be cast in the form
\begin{eqnarray}
12\left(\mu + A'^2\right)\left[\alpha A^2-2\beta\left(\mu + A'^2\right)\right]
-\Lambda A^4
= {12 \nu^2 \beta\over r^6}
\,\,\,.
\label{nu}
\end{eqnarray}
Our notation is explained in the footnote below\footnote{Since $A$ is a function of $|y|$
we have ${dA\over dy}=A'{d|y|\over dy}$ where $A'$ denotes the derivative 
of $A$ with respect to its argument $|y|$ and 
${d|y|\over dy}=2\Theta\left(y\right)-1$
where $\Theta\left(y\right)$ is the Heaviside function. Notice that $\left({d|y|\over dy}\right)^2=1$
and we have $\left({dA\over dy}\right)^2=A'^2$. On the other hand 
${d^2A\over dy^2}=A''+ 2A'\delta\left(y\right)$
where $A''$ denotes the second derivative of $A$ with respect to its argument $|y|$.}
. It is clear  
here that one must have $\nu=0$. This condition, however, is not needed if $\beta=0$ 
(see for example \cite{w2}).
With $\nu=0$,  the last equation takes then the simple form
\begin{eqnarray}
\mu + A'^2=\sigma A^2 
\,\,\,,
\label{A'}
\end{eqnarray}
where $\sigma$ is as previously defined.
The solution to our last differential equation is
\begin{eqnarray}
A\left(y\right)= {\mu\over 4\gamma\sigma}\exp\left(\pm \sqrt{\sigma}|y|\right)
+\gamma\exp\left(\mp\sqrt{\sigma}|y|\right)
\,\,\,.
\label{A}
\end{eqnarray}
Here $\gamma$ is an integration constant.
\par
We are left with one equation to solve, namely ${\cal H}^0_0$ 
\begin{eqnarray}
6\alpha A\left(\mu + A'^2\right) 
+6\left[\alpha A^2 -4\beta\left(\mu + A'^2\right)\right]{d^2A\over dy^2}
-\Lambda A^3 +
2\lambda A^3\delta\left(y\right) =0
\,\,\,.
\label{H00}
\end{eqnarray}
This equation 
involves second derivatives of $A\left(y\right)$ which generate delta functions
as explained in the footnote.
Using equation (\ref{A'})
in  ${\cal H}^0_0$ and matching the delta functions yields the fine tuning conditions
\begin{eqnarray}
\lambda = -6\left(\alpha -4\beta\sigma\right){A'(0)\over A(0)}
\,\,\,\,.
\label{lambda}
\end{eqnarray}
Substituting then for $A\left(y\right)$ in (\ref{H00}), fixes 
the five dimensional cosmological constant $\Lambda$ to its original value 
$\Lambda= 12\sigma\left(\alpha-2\beta\sigma\right)$.
Notice that the warp factor $A\left(y\right)$ can present various behaviours depending 
on the value of $\sigma$. In particular if $\sigma$ is negative then an oscillatory regime
is obtained.

\section{COSMOLOGICAL SOLUTIONS}

The metric for this study is taken to have the form
\begin{eqnarray}
ds^2=A^2(y)\left\{- dt^2+ a\left(t\right)^2\left[
{dr^2\over 1-kr^2}+r^2(d\theta^2+
\sin^2\theta d\phi^2)\right]\right\}+dy^2\,\,\,.
\end{eqnarray}
The non-vanishing components of the energy-momentum tensor $T^M_N$ are now given by
\begin{eqnarray}
T^0_0&=&-\rho\left(t\right)\delta(y) 
\nonumber\\
T^1_1&=&T^2_2=T^3_3=p\left(t\right)\delta(y)
\,\,\,. 
\end{eqnarray} 
The equations of motion we would like to solve are still ${\cal H}^M_N = {\cal E}^M_N+T^M_N=0$.
There are three different types of equations ${\cal H}^0_0$, 
${\cal H}^1_1={\cal H}^2_2={\cal H}^3_3$ and ${\cal H}^4_4$. The first of these, ${\cal H}^0_0$,
is 
\begin{eqnarray}
 6 \left[a^2\left(\alpha A^2 - 4\beta A'^2\right)
+4\beta\left(k+\dot{a}^2\right)\right]{d^2A\over dy^2}
-6\alpha A\left(k+\dot{a}^2\right)+6\alpha a^2AA'^2 -\Lambda A^3 a^2 
- 2\rho a^2A^3\delta\left(y\right)
=0
\,\,\,, 
\end{eqnarray} 
where $\dot{a}$ is the derivative of $a$ with respect to $t$.
Matching the delta functions in this last equation yields
\begin{eqnarray}
\rho = {6A'(0)\over a^2 \left[A(0)\right]^3}
\left[a^2\left(\alpha \left[A(0)\right]^2 -4 \beta \left[A'(0)\right]^2\right) 
+4\beta\left(k+\dot{a}^2\right)\right]
\,\,\,.
\end{eqnarray}
The second equation, ${\cal H}^1_1$, is given by the expression
\begin{eqnarray}
&& 2\left[3 a^2\left(\alpha A^2 -4 \beta  A'^2\right)  + 8\beta a\ddot{a} 
+4\beta\left(k+\dot{a}^2\right)\right]{d^2A\over dy^2}
-2\alpha A\left(k+\dot{a}^2\right) + 6\alpha a^2 A A'^2 - \Lambda a^2 A^3
-4\alpha a{\ddot{a}} A\nonumber\\
&&+ 2pa^2A^3\delta\left(y\right)=0
\,\,\,.
\end{eqnarray} 
Again, matching the delta functions coming from ${d^2A\over dy^2}$ and those coming from
the energy-momentum tensor gives
\begin{eqnarray}
p &=& -{2A'(0)\over a^2 \left[A(0)\right]^3}
\left[3 a^2\left(\alpha \left[A(0)\right]^2 -4 \beta \left[A'(0)\right]^2\right)   + 8\beta a\ddot{a} 
+4\beta\left(k+\dot{a}^2\right)\right]
\,\,\,.
\end{eqnarray}
Using the expression of $\rho$ we deduce the following expression for the Hubble parameter
\begin{eqnarray}
H^2\equiv \left({\dot{a}\over a}\right)^2 &=& {\left[A(0)\right]^3\over 24\beta A'(0)}\rho
-{k\over a^2} +\left(-{\alpha\left[A(0)\right]^2
\over 4\beta} +\left[A'(0)\right]^2\right)
\,\,\,.
\end{eqnarray}
This is the first Friedmann relation of ordinary cosmology. Similarly, combining
the expression of $\rho$ with that corresponding to $p$,
yields the second Friedmann equation
\begin{eqnarray}
\ddot{a} 
&=&  -{\left[A(0)\right]^3\over 48\beta A'(0)}\left(3p +\rho\right)a
 +\left(-{\alpha\left[A(0)\right]^2
\over 4\beta} +\left[A'(0)\right]^2\right)a
\,\,\,.
\end{eqnarray}
Differentiating $H^2$ and using the expression of $\ddot{a}$ results in the usual conservation equations
\begin{eqnarray}
&& \dot{\rho} a + 3\left(p +\rho\right)\dot{a}=0
\,\,\,.
\end{eqnarray}
Therefore, our gravitational theory with the Gauss-Bonnet term 
leads to ordinary Friedmann equations. 
\par 
Once the matching is carried out, we can deal with the two equations ${\cal H}^1_1$ and ${\cal H}^0_0$
away from $y=0$.  
Substracting ${\cal H}^1_1$ from ${\cal H}^0_0$ gives then
\begin{eqnarray}
\left(k+\dot{a}^2-a\ddot{a}\right)\left(\alpha A-4\beta A''\right)=0
\,\,\,\,.
\label{twopossible}
\end{eqnarray} 
If the first factor $\left(k+\dot{a}^2-a\ddot{a}\right)$ vanishes then an interesting solution
is given by
\begin{eqnarray}
a\left(t\right)={k\over 4\kappa\tau^2}\exp\left(\pm {\tau}t\right)
 +\kappa \exp\left(\mp {\tau}t\right)
\,\,\,\,,
\label{a}
\end{eqnarray} 
where $\tau$ and $\kappa$ are two integration constants.
\par
Upon substituting for $a(t)$ in ${\cal H}^4_4$, we obtain 
\begin{eqnarray}
12\left(A'^2 -{\tau^2}\right)\left[\alpha A^2 -2\beta\left(A'^2 -{\tau^2}\right)
\right] - \Lambda A^4=0
\,\,\,\,.
\end{eqnarray} 
This equation is exactly the one found in (\ref{nu}) with $\nu=0$ and 
where $\mu$ is replaced by $-{\tau^2}$. 
The solution, $A(y)$, to this equation is therefore as given in (\ref{A}) upon replacing
$\mu$ by $-{\tau^2}$. 
\par
Putting the expression of $a(t)$ in ${\cal H}^0_0$ leads to the differential equation
\begin{eqnarray}
6\alpha A\left(A'^2 -{\tau^2}\right) 
+6\left[\alpha A^2 -4\beta\left(A'^2 -{\tau^2}\right)\right]A''
-\Lambda A^3 =0
\,\,\,.
\end{eqnarray}
Again, this equation is that found in (\ref{H00}) away from the position of the  brane and 
where $\mu$ is replaced by $-{\tau^2}$. 
Therefore, replacing for $A(y)$ in this last differential equation 
fixes the bulk cosmological constant to be  
$\Lambda = 12 \sigma\left(\alpha -2\beta \sigma\right)$. 
\par
We should mention that this solution leads to the following relation
between the energy density $\rho$ and the pressure  $p$
\begin{eqnarray}
p= -\rho=\lambda
\,\,\,
\end{eqnarray}
where $\lambda$ is as given in (\ref{lambda}). The Hubble parameter for this solution
is given by
\begin{eqnarray}
H^2 =  -{k\over a^2} + \tau^2
\,\,\,
\end{eqnarray}
with $a\left(t\right)$ as given in (\ref{a}). 
Similarly, we find that
\begin{eqnarray}
{\ddot{a}\over a}=  \tau^2
\,\,\,
\end{eqnarray}
We distinguish, therefore, two cases. The first corresponds to $\tau^2>0$ and 
leads to an inflationary regime whenever one of the two exponentials in $a\left(t\right)$
dominates. The second
situation arises when  $\tau^2<0$. In this case the scale factor is given
by 
\begin{eqnarray}
a\left(t\right)= \varepsilon\cos\left({\sqrt{-k}\over\varepsilon}t+\varphi\right)
\,\,\,,
\end{eqnarray}
where $\varepsilon$ and $\varphi$ are two real integration constants. 
The corresponding $A\left(y\right)$ is given by equation (\ref{A}) where $\mu$ is replaced
by $-k/\varepsilon^2$. Of course, this solution is valid only when $k$ is negative and describes
a repeatedly collapsing universe. The Hubble parameter in this case is
$H^2 = -k/ a^2   + k/\varepsilon^2$.
\par 
The other solution to $\left(k+\dot{a}^2-a\ddot{a}\right)=0$ is given by
\begin{eqnarray}
a\left(t\right)=\pm\sqrt{-k} t + \delta
\,\,\,
\label{lineara}
\end{eqnarray}
with $\delta$ an integration constants. This solution is physical only
for $k$ negative. Substituting this expression of 
$a\left(t\right)$ in ${\cal{H}}^4_4$ leads to the differential
equation
\begin{eqnarray}
A'^2=\sigma A^2
\,\,\,,
\end{eqnarray}
where $\sigma$ is as defined in (\ref{sigma}). Therefore $A\left(y\right)$ is given by
\begin{eqnarray}
A\left(y\right)=\psi \exp\left(\sqrt{\sigma} |y|\right)\,\,\,\,\,\,\,\,\,\,{\rm {or}}
\,\,\,\,\,\,\,\,\,\,
A\left(y\right)= \omega \exp\left(-\sqrt{\sigma} |y|\right)
\,\,\,
\label{linearA}
\end{eqnarray}
for two integration constants $\psi$ and $\omega$. 
The solution in  (\ref{lineara}) and (\ref{linearA})  automatically satisfies ${\cal{H}}^0_0$.
Furthermore, we have $p= -\rho=\lambda$ and the Hubble parameter is $H^2 =-k/a^2$. 
\par
Let us now return to the second possibility as allowed by equation (\ref{twopossible}), namely when 
$\left(\alpha A-4\beta A''\right)=0$. This case is of course
present only when $\beta$ is different from zero and we have the solution
\begin{eqnarray}
A\left(y\right)=\xi \exp\left({1\over 2}\sqrt{{\alpha\over\beta}}|y|\right)
+ \theta \exp\left(-{1\over 2}\sqrt{{\alpha\over\beta}}|y|\right)
\,\,\,\,,
\label{secondA}
\end{eqnarray}
with $\xi$ and $\theta$ are two integration constants. Substituting this solution in ${\cal H}^4_4$ leads to
\begin{eqnarray}
&& a^3\left(3\alpha^2-2\beta\Lambda\right)
\left\{\left[\xi \exp\left({1\over 2}\sqrt{{\alpha\over\beta}}|y|\right)
+ \theta \exp\left(-{1\over 2}\sqrt{{\alpha\over\beta}}|y|\right)\right]^4
-6\xi^2\theta^2\right\}
\nonumber \\
&&-48\ddot{a}\left[\beta^2\left(k+\dot{a}^2\right)+\alpha\beta\xi\theta a^2\right]
-6\xi\theta a\left[8\alpha\beta\left(k+\dot{a}^2\right)
+\xi\theta a^2\left(5\alpha^2 +2\beta\Lambda\right)\right]=0
\,\,\,\,.
\end{eqnarray}
It is clear, from the separation of variables,  that one must have 
\begin{eqnarray}
\Lambda={3\alpha^2\over 2\beta}
\,\,\,\,
\end{eqnarray}
and the above differential equation reduces then to
\begin{eqnarray}
\left[\beta\left(k+\dot{a}^2\right)+\alpha\xi\theta a^2\right]
\left(\beta\ddot{a} + \alpha\xi\theta a\right)=0
\label{twofactors}
\,\,\,\,.
\end{eqnarray}
Regardless of which factor vanishes first, the solution to this last differential equation
takes the form
\begin{eqnarray}
a\left(t\right)=\zeta \exp\left(\sqrt{-{\alpha\xi\theta\over \beta}}t\right)
+ \chi \exp\left(-\sqrt{-{\alpha\xi\theta\over \beta}}t\right)
\,\,\,\,,
\label{seconda}
\end{eqnarray}
where $\zeta$ and $\chi$ are two integration constants and if the first factor
of the differential equation vanishes then 
\begin{eqnarray}
\beta k + 4\alpha\xi\theta\zeta\chi =0
\label{parameters}
\,\,\,\,.
\end{eqnarray}
With $\Lambda={\left(3\alpha^2\right)/\left(2\beta\right)}$, the equation corresponding to ${\cal H}^0_0$ 
yields 
\begin{eqnarray}
\left(\alpha A-4\beta A''\right)
\left[a^2\left(\alpha A^2 -4\beta A'^2\right) +4\beta\left(k+\dot{a}^2\right)\right]=0
\end{eqnarray}
and is therefore automatically satisfied.
\par
If the first factor in (\ref{twofactors}) vanishes then we have 
\begin{eqnarray}
p= -\rho =0
\,\,\,\,.
\end{eqnarray}
The Hubble parameter for this case is
\begin{eqnarray}
H^2 &=&  
-{k\over a^2}  -{\alpha\xi\theta\over\beta}
\,\,\,
\end{eqnarray}
On the other hand if the second factor in (\ref{twofactors}) vanishes then we have the 
equation of state
\begin{eqnarray}
p &=& -{1\over 3}\rho
=-8{A'(0)\over  \left[A(0)\right]^3}
{1\over a^2}
\left(\beta k +4\alpha\xi\theta\zeta\chi\right)
\,\,\,,
\label{eqnstate}
\end{eqnarray} 
where $A\left(y\right)$ is given by (\ref{secondA}).
The corresponding expression for the Hubble parameter is
\begin{eqnarray}
H^2 =
-{\alpha\xi\theta\over\beta}\left(1-{4\zeta\chi\over a^2}\right)
\,\,\,
\end{eqnarray} 
with $a(t)$ as given in (\ref{seconda}). In both cases we have
\begin{eqnarray}
{\ddot{a}\over a} =
-{\alpha\xi\theta\over\beta}
\,\,\,
\end{eqnarray}
We notice that if $-\alpha\xi\theta/\beta$ is positive then the universe is accelerating
otherwise $a\left(t\right)$ oscillates in time.

\section{Discussion}

Among the solutions found in this analyses, special attention should be paid to that
described by a time dependent energy density and pressure as in (\ref{eqnstate}). However, 
the equations of states $p=-\rho/3$ cannot correspond 
to an ordinary dust as the pressure $p$ is negative. 
Let us suppose that the only matter present on the brane is a self interacting scalar field with
energy-momentum tensor
\begin{eqnarray}
T_{ij} =\left[\partial_i\phi\partial_j\phi-{1\over 2}g_{ij}
\left(\partial^k\phi\partial_k\phi +V\left(\phi\right)\right)\right]\delta\left(y\right)
\,\,\,,
\end{eqnarray}
where $g_{ij}$ is a Friedmann-Robertson-Walker metric on the brane. 
If the field $\phi$ depends only on time then one has
\begin{eqnarray}
\rho &=&-T^0_0 = {1\over 2}\left[\dot{\phi}^2+V\right]\delta\left(y\right)
\nonumber\\
p &=& T^1_1 = T^2_2=T^3_3={1\over 2}\left[\dot{\phi}^2-V\right]
\delta\left(y\right)
\,\,\,.
\end{eqnarray} 
The field $\phi$ is subject to the equations of motion
\begin{eqnarray}
{1\over\sqrt{-g}}\partial_i\left(\sqrt{-g}\partial^i\phi\right)-{1\over 2}V'=0
\,\,\,
\end{eqnarray} 
with $V'$ being the derivative of the potential $V$ with respect to $\phi$. 
\par
The equation of state $p=-\rho/3$ leads to
\begin{eqnarray}
\dot\phi^2={1\over 2}V \,\,\,\,\,\,\,\,\,\Rightarrow\,\,\,\,\,\,\,\,\,
V'=4\ddot{\phi}
\,\,\,
\end{eqnarray} 
while the equation of motion of the scalar field yields
\begin{eqnarray}
\ddot\phi + 3{\dot a\over a}\dot\phi=-{1\over 2}V'
\,\,\,.
\end{eqnarray} 
Combining these last two equations results in
\begin{eqnarray}
{\ddot\phi\over \dot\phi} + {\dot a\over a}=0
\,\,\,.
\end{eqnarray}
Therefore $\dot\phi=\upsilon/a$, for some integration constant $\upsilon$.
Using the expression of $a\left(t\right)$ as given in (\ref{seconda}), we find that
\begin{eqnarray}
\phi\left(t\right)=
{\upsilon\over\sqrt{-{\alpha\xi\theta\zeta\chi\over \beta}}}
\arctan\left[\sqrt{{\zeta\over\chi}}\exp\left(\sqrt{-{\alpha\xi\theta\over \beta}}t\right)
\right]+\varpi \,\,\,\,,
\end{eqnarray}
where $\varpi$ is an integration constant. In order to be able to express the potential $V$
as a fonction of $\phi$, it is convenient to rewrite this last equation in the form  
\begin{eqnarray}
\exp\left(\sqrt{-{\alpha\xi\theta\over \beta}}t\right)=
\sqrt{{\chi\over\zeta}}\tan\left[{1\over\upsilon}\sqrt{-{\alpha\xi\theta\zeta\chi\over \beta}}
\left(\phi-\varpi\right)\right]
\,\,\,.
\end{eqnarray}
Since the potential is given by 
$V=2\dot\phi^2=2\upsilon^2/a^2$, we find that
\begin{eqnarray}
V\left(\phi\right)={\upsilon^2\over 2\zeta\chi}\sin^2
\left[{2\over\upsilon}\sqrt{-{\alpha\xi\theta\zeta\chi\over \beta}}
\left(\phi-\varpi\right)\right]
\,\,\,.
\end{eqnarray}
This potential has an infinite number of minima.
\par
As it is known, cosmological scenarios with negative pressure are used
in explaining the current accelaration of our universe (see for example \cite{supernova}).
The scalar field providing this pressure is known as quintessence \cite{peebles}.  Similar models are 
also constructed in the context of brane world cosmology \cite{quin1,quin2,quin3,quin4}. 
It seems that we have found here another model for quintessence where the scalar field is governed by a 
very simple potential.

\smallskip
\smallskip
\smallskip

\noindent $\underline{\hbox{\bf Acknowledgments}}$: One of us (BA) would like 
to thank Professor Peter Forg\'acs for his kind invitation to the University of Tours where  
this work was carried out. He is also grateful to the members of the group 
for their kind hospitality. (NM) would like to thank Claude Barrabes, Bruno Boisseau, Bernard Linet
and Micheal Volkov for very useful discussions.

\end{document}